\documentclass[runningheads,a4paper,11pt]{llncs}
\usepackage[text={150mm,247mm},centering]{geometry} %% 210 mm x 297 mm

\usepackage{multirow}
\usepackage{amsmath, amssymb}
\usepackage[english]{babel}
\usepackage{graphicx}
\usepackage{xspace, url, verbatim, array, float}
\usepackage[mathscr]{eucal}
\usepackage{cite}
\usepackage{comment}

\usepackage{dcolumn}
\newcolumntype{S}{D{.}{.}{5}} %% for masses

\newcommand\abs[1]{\left\lvert#1\right\rvert}

\newcommand\tte{\textsc{Ten~times~18}\xspace}

\newcommand\Prob{\mathbb{P}}
\newcommand\Exp{\mathbb{E}}
\newcommand\xmin{x_{\min}}
\newcommand\xmax{x_{\max}}

\begin{document}

\title{Ten times eighteen}

\titlerunning{Ten times eighteen}

\author{Sebastian B\"ocker}

\authorrunning{S.~B\"ocker}

\institute{Chair for Bioinformatics, Friedrich-Schiller-University, Jena,
  Germany, \url{sebastian.boecker@uni-jena.de}}

\date{\today}

\maketitle

%%%%%%%%%%%%%%%%%%%%%%%%%%%%%%%%%%%%%%%%%%%%%%%%%%%%%%%%%%%%%%%%%%%%%%%%%%%%
%%%%  ABSTRACT  %%%%%%%%%%%%%%%%%%%%%%%%%%%%%%%%%%%%%%%%%%%%%%%%%%%%%%%%%%%%
%%%%%%%%%%%%%%%%%%%%%%%%%%%%%%%%%%%%%%%%%%%%%%%%%%%%%%%%%%%%%%%%%%%%%%%%%%%%

\begin{abstract}
  We consider the following simple game: We are given a table with ten slots
  indexed one to ten.  In each of the ten rounds of the game, three dice are
  rolled and the numbers are added.  We then put this number into any free
  slot.  For each slot, we multiply the slot index with the number in this
  slot, and add up the products.  The goal of the game is to maximize this
  score.  In more detail, we play the game many times, and try to maximize
  the sum of scores or, equivalently, the expected score.  We present a
  strategy to optimally play this game with respect to the expected score.
  We then modify our strategy so that we need only polynomial time and space.
  Finally, we show that knowing all ten rolls in advance, results in a
  relatively small increase in score.  Although the game has a random
  component and requires a non-trivial strategy to be solved optimally, this
  strategy needs only polynomial time and space.
\end{abstract}

%%%%%%%%%%%%%%%%%%%%%%%%%%%%%%%%%%%%%%%%%%%%%%%%%%%%%%%%%%%%%%%%%%%%%%%%%%%%%
%%%%  SECTION  %%%%%%%%%%%%%%%%%%%%%%%%%%%%%%%%%%%%%%%%%%%%%%%%%%%%%%%%%%%%%%
%%%%%%%%%%%%%%%%%%%%%%%%%%%%%%%%%%%%%%%%%%%%%%%%%%%%%%%%%%%%%%%%%%%%%%%%%%%%%

\section{Introduction}

\begin{figure}[b]
\centering
\includegraphics[width=0.9\textwidth]{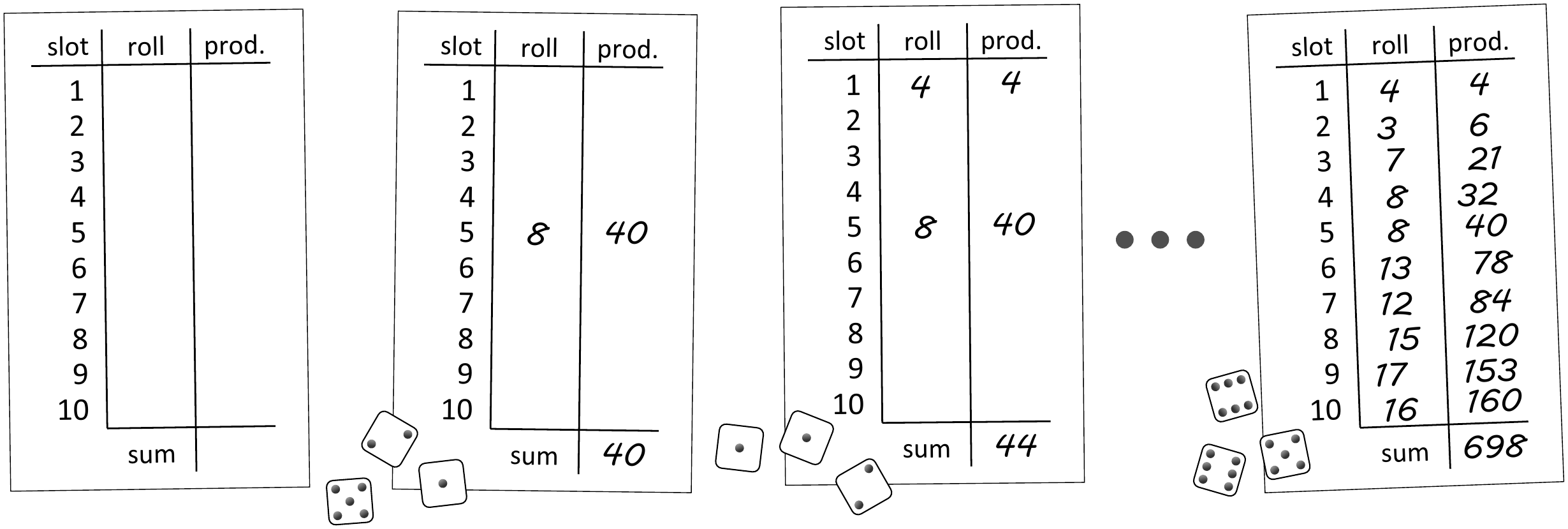}
\caption{Example of a \tte game.  The first roll is `8', and the player
  chooses to place it in slot \#5, resulting in product score~$40$.  The
  second roll is `4', and the player chooses to place it in slot \#1,
  resulting in product score~$4$ and sum of products~$44$.  At the end of the
  game, the player has reached total score~$698$, an excellent score as we
  will see below, compare to Table~\ref{tab:important-scores}.  Note that due
  to incomplete information, the player has made several suboptimal choices.}
\label{fig:example-game}
\end{figure}

When I was in twelfth grade at school, my computer science teacher introduced
us to the following game: Assume that you are given a table with ten slots
indexed one to ten.  The game proceeds in ten rounds.  In each round, three
dice are rolled and the numbers are added.  Then, you are allowed to put this
number into any free slot.  In the end, your table is completely filled with
numbers between three and~$18$.  For each slot, you multiply the slot index
with the number in this slot, and then you add up the products.  An example
is given in Fig.~\ref{fig:example-game}.  The goal of the \tte game is to
maximize the sum of products.  The smallest total score that you can reach is
\[
  (1 + 2 + \dots + 9 + 10) \cdot 3 = 165,
\]
the largest score is~$990$.

%%%%%%%%%%%%%%%%%%%%%%%%%%%%%%%%%%%%%%%%%%%%%%%%%%%%%%%%%%%%%%%%%%%%%%%%%%%%%
%%%%%%%%%%%%%%%%%%%%%%%%%%%%%%%%%%%%%%%%%%%%%%%%%%%%%%%%%%%%%%%%%%%%%%%%%%%%%

If you play \tte, you will quickly come up with first ideas whether
certain moves that are good or bad: For example, you should definitely put a
``three'' into slot number one, and you should put an eighteen into slot
number ten.  If these slots are not available, put them in the slot with the
smallest or highest index available, respectively.  But what about a roll of
``five''?  And what do you do if you roll a ``seven'' and all even slots have
been taken?  Is this basically the same problem as rolling a seven when all
odd slots have been taken?  (In fact, it is.)

The question this boils down to, is: How do we maximize the sum of products?
That is, we are searching for a \emph{strategy} that maximizes our chances of
winning, that is, the points we can obtain.  Clearly, playing only a single
game is not sufficient to judge a strategy, so we repeat the game many times
and for all these games, we again sum up the sum of products.  Formally
speaking, this boils down to: What is a strategy that maximizes the
\emph{expected} sum of products?

Back in 1987, my schoolmates and I came up with many different strategies for
\tte: These were based on statistical considerations, and even some simple
machine learning strategies (play the game repeatedly and see what moves are
favorable).  Funnily, it is rather straightforward to find an optimal
strategy if you are familiar with the concept of dynamic programming --- and
a tiny twist.  In fact, you do not need a fancy computer to find this
strategy.  And with a little more statistics, we can even find a strategy
that optimally plays practically any variant of \tte: That is, the strategy
requires only polynomial time and space.  Finally, we show how to compute the
advantage of an ``all-knowing'' strategy, which is allowed to look into the
future before placing the rolls: Interestingly, this advantage is relatively
small.

Playing \tte is different from many other solitaire games in that rolling
dice is involved.  Combinatorial games without chance (such as Rubik's cube)
have been studied more frequently~\cite{berlekamp01winning}, in particular
the complexity of playing an optimal strategy.  The probably ``closest
relative'' to \tte is Yahtzee, a popular dice game.  Optimal solitaire
strategies --- again in the sense of maximizing the expected score --- were
independently developed by Tom Verhoeff~\cite{verhoeff10solitaire} and James
Glenn~\cite{glenn06optimal} around 1999, but never formally published.  To
this end, further authors developed optimal strategies for the solitaire
game~\cite{vancura01advantage, woodward03yahtzee}.  Obviously, Yahtzee is
much more involved than \tte, and so is the analysis of the game.

%%%%%%%%%%%%%%%%%%%%%%%%%%%%%%%%%%%%%%%%%%%%%%%%%%%%%%%%%%%%%%%%%%%%%%%%%%%%%
%%%%  SECTION  %%%%%%%%%%%%%%%%%%%%%%%%%%%%%%%%%%%%%%%%%%%%%%%%%%%%%%%%%%%%%%
%%%%%%%%%%%%%%%%%%%%%%%%%%%%%%%%%%%%%%%%%%%%%%%%%%%%%%%%%%%%%%%%%%%%%%%%%%%%%

\section{Preliminaries}

Let $B := \{1,\dots,10\}$ be the slots, and let $A \subseteq B$ be the slots
that have already been filled.  The first important thing to notice, is that
for finding the best move at this point, it does not matter what numbers have
actually been inserted into the slots that have been filled: You can simply
think of it as a new game where an incomplete table has been given to you,
and your task is to maximize the sum of products for the incomplete table.
Doing so, we also maximize the sum of products for the complete table.  The
score of the complete table obviously depends on the previously filled slots;
but we cannot change a previous decision.

We now formalize the problem a little bit: We model rolling the three dice as
a \emph{random variable} $X$ with $X \in \{3,\dots,18\}$.  We denote the
probability that some happens by $\Prob(\text{event})$.  The probability that
we roll a three is $1$ in $216$, which is formally written as $\Prob(X=3) =
\frac{1}{216}$.  Similarly, we are given the probabilities $\Prob(X=4) =
\frac{3}{216}$ and so on, see Table~\ref{tab:probs}.  We assume that $X$ is
always an integer, and that there exist integer bounds $\xmin,\xmax$ such
that $\xmin \le X \le \xmax$.  Using this formal variable, allows us to
re-use our thoughts below for other variants of the \tte game: For example,
the dices might be loaded; we might want to throw two or four dices instead
of three; or, we might even throw five twelve-sided dice.  For all of these
variants, the solution introduced below works, though you have to repeat the
calculations.

%%%%%%%%%%%%%%%%%%%%%%%%%%%%%%%%%%%%%%%%%%%%%%%%%%%%%%%%%%%%%%%%%%%%%%%%%%%%%
\begin{table}[b]
\centering
\begin{tabular}{r|rrrrrrrrrrrrrrrr}
  $x$ & 3 & 4 & 5 & 6 & 7 & 8 & 9 & 10 & 11 & 12 & 13 & 14 & 15 & 16 & 17 &
  18 \\

  \hline

  $216 \cdot \Prob(X=x)$ & 1 & 3 & 6 & 10 & 15 & 21 & 25 & 27 & 27 & 25 & 21
  & 15 & 10 & 6 & 3 & 1
\end{tabular}

\medskip
\caption{Probabilities for throwing three dice.}
\label{tab:probs}
\end{table}
%%%%%%%%%%%%%%%%%%%%%%%%%%%%%%%%%%%%%%%%%%%%%%%%%%%%%%%%%%%%%%%%%%%%%%%%%%%%%

For a given random variable $Y$ we denote its expected value as $\Exp(Y)$.
When the probabilities of all possible outcomes are known to us, we can
compute the expected value by summing over the products of the probability
times the outcome.  For the three dice example with random variable $X$ we
can calculate
\[
  \Exp (X) = \frac{1}{216} \cdot 3 + \frac{3}{216} \cdot 4 + \dots +
  \frac{1}{216} \cdot 18 = 10.5 .
\]
Clearly, there is a simpler way to calculate this: For two random variables
$X,Y$ we have $\Exp(X+Y) = \Exp(X) + \Exp(Y)$.  In other words, the expected
value of the sum of three identical dices equals three times the expected
value of a single dice.  If we assume that $X'$ is the random variable of a
single dice, then $\Exp(X) = 3 \, \Exp(X') = 3 \cdot 3.5 = 10.5$.

The simplest strategy that we can evaluate using the above considerations, is
the ``random strategy'' where we assign each roll randomly to a slot.  This
strategy has expected score
\[
  (1 + 2 + \dots + 9 + 10) \cdot 10.5 = 577.5 .
\]
This score is what we have to compare our strategy against in the future.

%%%%%%%%%%%%%%%%%%%%%%%%%%%%%%%%%%%%%%%%%%%%%%%%%%%%%%%%%%%%%%%%%%%%%%%%%%%%%
%%%%  SECTION  %%%%%%%%%%%%%%%%%%%%%%%%%%%%%%%%%%%%%%%%%%%%%%%%%%%%%%%%%%%%%%
%%%%%%%%%%%%%%%%%%%%%%%%%%%%%%%%%%%%%%%%%%%%%%%%%%%%%%%%%%%%%%%%%%%%%%%%%%%%%

\section{Why is this complicated?}

Often, people who get to know \tte immediately start thinking about one or
the other strategy to solve it.  One particular, general approach easily
comes into mind: Why not model the complete game as one large decision tree
where nodes correspond to states of what has happened so far, and edges
correspond to changing from one state to another?  That is, we start with an
initial state where all slots are empty.  Then, we add 160 outgoing edges,
one for each roll from 3 to 18 and one for each slot that we can fill with
it.  In the end, we will only have to store the optimal slot to be filled
with each number; but as we are only in the process of determining this
optimal slot, storing the complete tree appears to be
inevitable.\footnote{Similar trees are used for many games, in particular
  two-player games with complete knowledge and without chance, such as
  chess.}

One can easily check that this approach suffers from the size of the tree
that we have to compute and store: As noted above, there are $16 \cdot 10 =
160$ outgoing edges from the root node, resulting in the same number of nodes
in the tree.  Leaving every such node, there are $16 \cdot 9 = 144$ outgoing
edges and a total of $160 \cdot 144 = 23\,040$ nodes at the next level.  In
total, we reach
\[
  16^{10} \cdot 10! = 16 \cdot \dots \cdot 16 \cdot 10 \cdot 9 \cdot \dots
  \cdot 2 \cdot 1 = 3.99 \cdot 10^{18}
\]
at the last level of the tree.  So, storing some value for each node of the
tree is impossible on today's computers, and even beyond the capacity of any
hard disk, as it requires several exabytes of memory.  Hence, this road is
blocked, in particular if you want to play \tte with more than ten slots, see
below.

From a computational complexity viewpoint, the arguably most interesting
question is: can we decide with polynomial time and space upon the next
optimal move, or is the problem NP-hard~\cite{garey79computers}?  For a
polynomial algorithm, we require that time and space are bounded by a
polynomial in all aspects of the input: the number of highest roll and, in
particular, the number of slots in the input.  We will come back to this
question in Sec.~\ref{sec:polynomial}.

%%%%%%%%%%%%%%%%%%%%%%%%%%%%%%%%%%%%%%%%%%%%%%%%%%%%%%%%%%%%%%%%%%%%%%%%%%%%%
%%%%  SECTION  %%%%%%%%%%%%%%%%%%%%%%%%%%%%%%%%%%%%%%%%%%%%%%%%%%%%%%%%%%%%%%
%%%%%%%%%%%%%%%%%%%%%%%%%%%%%%%%%%%%%%%%%%%%%%%%%%%%%%%%%%%%%%%%%%%%%%%%%%%%%

\section{Dynamic Programming}

Dynamic programming solves complex problems by breaking them down into
simpler subproblems.  To solve a problem, we need to solve different parts of
the problem (subproblems), then combine the solutions of the subproblems to
reach an overall solution.  We make sure that each subproblem is solved only
once, thus reducing the number of computations.  Top-down dynamic programming
simply means storing the results for all subproblems that we encounter.  In
bottom-up dynamic programming, we try to solve smaller subproblems first, and
deduce the solution of larger subproblems by combining those of smaller
subproblems.  We will concentrate on bottom-up dynamic programming, so that
our solution does not require any recursive calls.

Assume that slots $A \subseteq B := \{1,\dots,10\}$ have been filled before.
We want to know what score we can reach for the rest of the game, if we play
an ``optimal strategy''.  This optimality depends on the rolls that will
happen in the future, so we cannot talk about \emph{the} score that we will
obtain.  But what we can do is to talk about the expected value of the score
that we can reach; it is this score that we want to maximize.  To this end,
let $M[A]$ denote the maximum expected value of the score that we can reach
using any strategy.  Then, $M[B]$ is the maximum expected score that we can
reach for the complete game.  In fact, we are rather interested in the
strategy that leads to this maximum expected score, and not so much in the
score itself.  But as so often in dynamic programming, let us forget about
the structure of the solution (that is, the strategy) for the moment and
concentrate solely on its score.  As it will turn out, it is rather simple to
come up with the strategy as soon as the matrix has been filled.

There exist $2^{\abs{B}}$ subsets of the set $B$, including the empty set and
the full set.  This comes down to $2^{10} = 1024$ subsets for $B =
\{1,\dots,10\}$.  For each subset $A \subseteq B$ we store the entry $M[A]$.
In implementation, the subsets $A$ will be represented as bit vectors, and
every subset $A$ can be easily transformed into a number between $0$ and
$2^{\abs{B}} -1$.

We have noted above that one trick of dynamic programming is to compute the
solutions for each subproblem only once, and to store it so it can be
accessed multiple times.  Here, this means that we want to compute the
entries of table $M$ in the right order, and to use previously computed
entries of $M$ for deriving the next one.  In particular, we want to make
sure that any entry of the matrix $M$ is accessed only after it has been
computed.  To this end, we first need an initialization to start from: If
none of the slots has been filled so far, then the best expected score is
obviously zero for doing nothing, so $M[\emptyset] = 0$.  It is a well-known
trick to initialize the dynamic programming table for an entry where, in
fact, nothing has happened so far.  If you do not like the empty set
initialization, you can instead initialize
\[
  M \bigl[ \{i\} \bigr] = i \cdot \Exp(X) \quad \text{for $i =
    \xmin,\dots,\xmax$}
\]
because moving any number to the last remaining slot $i$, the expected score
for doing so is simply $i \cdot \Exp(X)$.  This initialization is slightly
more complicated but leads to exactly the same results.

To make sure that we only access entries of the table that have been
previously computed, we iterate $k = 1,\dots,\abs{B}$, and in each step of
the iteration we compute all entries $M[A]$ for all subsets $A \subseteq B$
with $\abs{A} = k$.  (If you have initialized the one-element subsets you can
leave out $k=1$ in the iteration.)  To this end, assume that the table $M$
has been filled for all $A \subseteq B$ where $\abs{A} \le k-1$.  We now show
how to compute it for each entry $A \subseteq B$ with $\abs{A} = k$.  This
means that we are allowed to distribute $k$ rolls into the filled slots~$A$.
We concentrate on the next roll: The probability that some $x$ with $\xmin
\le x \le \xmax$ is rolled next, is $\Prob(X=x)$.  Possible rolls are lower
bounded by $\xmin$ and upper bounded by~$\xmax$.  If we decide to put roll
$x$ into slot $i$ for $i \in A$ then we gain $i \cdot x$ in the sum of
products.  Playing the remaining slots, the best strategy will (by definition
of~$M$) reach expected score $M[A - \{i\}]$.  Putting this together we get
\begin{equation} \label{equ:recurrence}
  M[A] = \sum_{x = \xmin,\dots,\xmax} \Prob (X = x) \cdot \max_{i \in A}
  \Bigl\{ i \cdot x + M[A - \{i\}] \Bigr\}
\end{equation}

How long does it take to fill the matrix~$M$?  There exist $2^{\abs{B}}$ many
entries in the matrix.  For each entry we iterate over $\xmax-\xmin+1$ many
values for~$x$, and $\abs{A} \le \abs{B}$ different values for~$i$, a total
of $O \bigl( (\xmax-\xmin+1) \cdot \abs{B} \bigr)$ entries.\footnote{The
  ``big~O'' notation is used to describe the asymptotic behavior of some
  function, ignoring constant factors.}  In total, filling the complete
matrix requires $O \bigl( 2^{\abs{B}} \cdot (\xmax-\xmin+1) \cdot \abs{B}
\bigr)$ time: That is, we need less than $c \cdot 2^{\abs{B}} \cdot
(\xmax-\xmin+1) \cdot \abs{B}$ summations, multiplications, and comparisons
for some multiplicative constant~$c$.

Now, the maximum expected score that any strategy can reach, can be computed
as
\[
  M[\{1,\dots,10\}] = 642.2393504256 .
\]
This score can be computed using those entries $M[A]$ where $A$ has
cardinality~$9$, see Table~\ref{tab:M-nine}.  Due to space constraints, we
cannot show all 1024 entries of the table.  Note that the expected score
drops with higher $i$: This is as we would expect it, because for small $i$
we have already used up more of the high-scoring slots.

%% 642.2393504256
%% 618.3200125344
%% 611.4500004391
%% 603.3935460429
%% 594.4280942260
%% 584.6684156968
%% 574.1684156968
%% 562.9280942260
%% 550.8935460429
%% 537.9500004391
%% 523.8200125344

%%%%%%%%%%%%%%%%%%%%%%%%%%%%%%%%%%%%%%%%%%%%%%%%%%%%%%%%%%%%%%%%%%%%%%%%%%%%%
\begin{table}[tb]
\centering
\begin{tabular}{l|ccccc}
$i$ & 1 & 2 & 3 & 4 & 5 \\

\hline

$M[B-\{i\}]$ & 618.32001 & 611.45000 & 603.39355 & 594.42809 & 584.66842
\end{tabular}

\medskip

\begin{tabular}{l|ccccc}
$i$ & 6 & 7 & 8 & 9 & 10 \\

\hline

$M[B-\{i\}]$ & 574.16842 & 562.92809 & 550.89355 & 537.95000 & 523.82001
\end{tabular}

\medskip
\caption{The matrix $M$ for all subsets $A \subseteq B = \{1,\dots,10\}$ of
  cardinality~$9$, rounded to five decimal places.  This table is required to
  decide upon the first move of \tte.}
\label{tab:M-nine}
\end{table}
%%%%%%%%%%%%%%%%%%%%%%%%%%%%%%%%%%%%%%%%%%%%%%%%%%%%%%%%%%%%%%%%%%%%%%%%%%%%%

%%%%%%%%%%%%%%%%%%%%%%%%%%%%%%%%%%%%%%%%%%%%%%%%%%%%%%%%%%%%%%%%%%%%%%%%%%%%%
%%%%  SECTION  %%%%%%%%%%%%%%%%%%%%%%%%%%%%%%%%%%%%%%%%%%%%%%%%%%%%%%%%%%%%%%
%%%%%%%%%%%%%%%%%%%%%%%%%%%%%%%%%%%%%%%%%%%%%%%%%%%%%%%%%%%%%%%%%%%%%%%%%%%%%

\section{Playing the game}

How does knowledge about the maximum expected score, $M[A]$, help us to come
up with a useful move?  This, in fact, is quite simple: Assume that slots $A
\subseteq B$ have previously been filled, and that number $x$ has been rolled
in this move.  From the above, it is straightforward to show that the maximum
expected score that we can reach after we have placed $x$ is
\begin{equation}
  \max_{i \in B - A} \Bigl\{ i \cdot x + M[A \cup \{i\}] \Bigr\} .
\end{equation}
This follows because $M[A \cup \{i\}]$ is the maximum expected value that we
can reach when slot $A \cup \{i\}$ have been filled previously.  So, all we
have to do is search for $i^*$ such that
\begin{equation} \label{equ:best-choice}
  i^* \cdot x + M \bigl[ A \cup \{i^*\} \bigr] = \max_{i \in B - A} \Bigl\{ i
  \cdot x + M[A \cup \{i\}] \Bigr\}
\end{equation}
and then, place $x$ in slot~$i^*$.  This can be achieved quickly: We need
only $O(\abs{B})$ steps to find the maximum.

%% Place 3 in slot 1 with score 621.3200125344
%% Place 4 in slot 1 with score 622.3200125344
%% Place 5 in slot 1 with score 623.3200125344
%% Place 6 in slot 1 with score 624.3200125344
%% Place 7 in slot 2 with score 625.4500004391
%% Place 8 in slot 2 with score 627.4500004391
%% Place 9 in slot 4 with score 630.4280942260
%% Place 10 in slot 5 with score 634.668415696
%% Place 11 in slot 6 with score 640.168415696
%% Place 12 in slot 7 with score 646.928094226
%% Place 13 in slot 9 with score 654.950000439
%% Place 14 in slot 9 with score 663.950000439
%% Place 15 in slot 10 with score 673.820012534
%% Place 16 in slot 10 with score 683.820012534
%% Place 17 in slot 10 with score 693.820012534
%% Place 18 in slot 10 with score 703.820012534

%%%%%%%%%%%%%%%%%%%%%%%%%%%%%%%%%%%%%%%%%%%%%%%%%%%%%%%%%%%%%%%%%%%%%%%%%%%%%
\begin{table}[tb]
\centering
\begin{tabular}{l|cr}
roll & slot & $\Exp(\text{score})$ \\
\hline
3 & \#1 & 621.32001 \\
4 & \#1 & 622.32001 \\
5 & \#1 & 623.32001 \\
6 & \#1 & 624.32001
\end{tabular}
\quad
\begin{tabular}{l|cr}
roll & slot & $\Exp(\text{score})$ \\
\hline
7 & \#2 & 625.45000 \\
8 & \#2 & 627.45000 \\
9 & \#4 & 630.42809 \\
10 & \#5 & 634.66842
\end{tabular}
\quad
\begin{tabular}{l|cr}
roll & slot & $\Exp(\text{score})$ \\
\hline
11 & \#6 & 640.16842 \\
12 & \#7 & 646.92809 \\
13 & \#9 & 654.95000 \\
14 & \#9 & 663.95000
\end{tabular}
\quad
\begin{tabular}{l|cr}
roll & slot & $\Exp(\text{score})$ \\
\hline
15 & \#10 & 673.82001 \\
16 & \#10 & 683.82001 \\
17 & \#10 & 693.82001 \\
18 & \#10 & 703.82001
\end{tabular}

\medskip
\caption{The best strategy for the first move of \tte.  For each roll, slot
  index $i^*$ has been chosen using eq.~\eqref{equ:best-choice}.}
\label{tab:first-move}
\end{table}
%%%%%%%%%%%%%%%%%%%%%%%%%%%%%%%%%%%%%%%%%%%%%%%%%%%%%%%%%%%%%%%%%%%%%%%%%%%%%

We have depicted the ``maximum expected score'' strategy for the first move
of the game in Table~\ref{tab:first-move}, including the expected score that
we can reach including this first move.  There are at least two unexpected
things to notice in this table: Firstly, even a roll of $6$ should still be
placed in the first slot, and similarly, even a roll of $15$ should still be
placed in the highest slot.  This becomes understandable, though, if we
consider that rolling a $3$ to $6$ has total probability of less than
$10\,\%$; and the same holds for rolling a $15$ to~$18$.  Second, it never
pays off to put the first roll into slots \#3 or \#8.  It is doubtful that
there is a simple explanation for this fact; it simply comes out of our
calculations.

We can also ask for the ``closest call'' of the ``maximum expected score''
strategy: In which move are two different slots the closest in the difference
of expected score we will reach after this placement?  For the first move,
this is a roll of $9$: If we place it into slot \#3 (instead of the optimal
slot \#4) we can still reach an expected score of $630.39355$, the difference
being only $0.03455$.  Similarly, we can place a roll of $12$ into slot \#8
instead of slot \#7, with the same difference in score.  For the complete
game, the closest call is placing a roll of $10$ when seven consecutive slots
are available: Here, the runner-up placement of the roll decreases the
expected score by $0.02989$.

%%%%%%%%%%%%%%%%%%%%%%%%%%%%%%%%%%%%%%%%%%%%%%%%%%%%%%%%%%%%%%%%%%%%%%%%%%%%%
%%%%  SECTION  %%%%%%%%%%%%%%%%%%%%%%%%%%%%%%%%%%%%%%%%%%%%%%%%%%%%%%%%%%%%%%
%%%%%%%%%%%%%%%%%%%%%%%%%%%%%%%%%%%%%%%%%%%%%%%%%%%%%%%%%%%%%%%%%%%%%%%%%%%%%

\section{Polynomial time and space} \label{sec:polynomial}

The above ``maximum expected score'' strategy requires us to compute and
store an array with $2^{\abs{B}}$ entries.  This is not a problem for
$\abs{B} = 10$, as the total size of the table is only $1024$.  Even in the
1990s, practically every home computer came with a sufficient amount of
memory to store such a table.\footnote{The only notable exception that I am
  aware of was the Sinclair ZX81 where the basic model shipped with only
  1~kilobyte of Random Access Memory.}  But the important point is that
memory requirement increases \emph{exponentially} with the size of the
set~$B$.  Whereas one could think of the analogous games with $\abs{B} = 20,
30, 40$ as being twice (three times or four times, respectively) as hard as
the original game, we need megabytes, gigabytes, or even terabytes to store
the table~$M$.  This implies that for $\abs{B} = 40$ tables are already much
to large to be stored in the main memory of the average present-day
computers.  Given that the current rate of miniaturization integrated
circuits is kept throughout the next years, it would still require more than
a year so that we can increase the size of solvable instances by \emph{one}.
Even if every atom in the observable universe (approximately $10^{80}$) would
be used to store one entry of our table~$M$, this would not allow us to play
a game where $\abs{B} > 265$.  Therefor, it is an interesting question
whether we can get away with less memory.

We can answer this question easily for one particular type of \tte: That
is, if we have only two possible outcomes for each throw (flipping a coin),
such as $1$ and~$2$.  In this case, the problem becomes trivial: Just place
any $1$ into the first available slot (with smallest index), and place any
$2$ into the last available slot (with highest index).  It is clear that this
strategy reaches the optimum expected score, uses constant memory and
performs each move in constant time.

But somewhat unexpectedly, we can still find a solution for our original game
(and, in fact, any variant of \tte where slot multipliers are strictly
increasing).  Unfortunately, we need a little more statistics to show that we
can actually solve the problem with polynomial memory and time.  Assume that
there are $k$ slots left, and that
\[
  0 < \lambda_1 < \lambda_2 < \dots < \lambda_k
\]
are the score multipliers.  For any deterministic or random strategy, let
$Y_1,\dots,Y_k$ be random variables such that $Y_i$ is the roll the
strategy places on slot~$i$.  Now, $Y := \lambda_1 Y_1 + \dots + \lambda_k
Y_k$ is the random variable for the score of this strategy, and we have
\begin{equation} \label{equ:exp-score}
  \Exp(Y) = \Exp \bigl( \lambda_1 Y_1 + \dots + \lambda_k Y_k \bigr) =
  \lambda_1 \Exp(Y_1) + \dots + \lambda_k \Exp(Y_k) .
\end{equation}
Note that the random variables $Y_i$ are strongly correlated, as placing a
roll of $18$ into the highest slot will influence the expected values for all
other slots; but \eqref{equ:exp-score} also holds for correlated random
variables.  Assume that there exist $i < j$ such that $\Exp(Y_i) >
\Exp(Y_j)$.  Then, the strategy cannot be optimal: simply exchange all moves
of the strategy to slots~$i$ and~$j$, what results in a strategy with
expected score
\[
  \Exp(Y) + \bigl( \Exp(Y_i) - \Exp(Y_j) \bigr) \cdot (\lambda_j - \lambda_i)
  > \Exp(Y)
\]
as $\lambda_j > \lambda_i$ and, by our assumption, $\Exp(Y_i) \ge \Exp(Y_j)$.
This implies that for an optimal strategy, we have
\begin{equation} \label{equ:exp-ordered}
  \Exp(Y_1) \le \Exp(Y_2) \le \dots \le \Exp(Y_k) .
\end{equation}

Assume that $k+1$ slots are empty, and that our roll is some $x \in
\{\xmin,\dots,\xmax\}$ --- where will the best strategy to maximize the
expected score place this roll?  From \eqref{equ:exp-ordered} it is
straightforward to understand that this roll must be placed on the $i$-th
free slot such that $\Exp(Y_{i-1}) \le x \le \Exp(Y_i)$: We can easily show
that placing $x$ into any other free slot, will result in a suboptimal
expected score.  This means that the $\lambda_i$ are not taken into
consideration for deciding upon the best move.

To this end, let us consider the ``maximum expected score'' strategy for $j$
empty slots with strictly increasing slot weights: We define $E_j[i] =
\Exp(Y_i)$ as the expected value of the $i$-th slot.  It is easy to
understand how $E_{j+1}[\cdot]$ can be computed from $E_j[\cdot]$: For
$E_j[\cdot]$ and $x \in \{\xmin,\dots,\xmax\}$ we define $I_{j}(x)$ as the index such
that
\[
  E_{j}[i-1] \le x \le E_j[i] \quad \text{for $i = I_j(x)$} .
\]
We may assume that $E_j[0] = -\infty$ and $E_j[j+1] = +\infty$. In case of a
draw we can choose any such index.  We infer the recurrence:
\begin{equation}
  E_j[i] = \sum_{x = \xmin,\dots,\xmax} \Prob(X = x) \cdot
    \begin{cases}
      E_{j-1}[i] & \text{for $i = 1,\dots,I_j(x)-1$} \\
      x & \text{for $i = I_j(x)$} \\
      E_{j-1}[i-1] & \text{for $i = I_j(x)+1,\dots,j$}
    \end{cases}
\end{equation}
In the end, the expected score of the ``maximum expected score'' strategy can
be calculated as $\sum_{i=1}^k i \cdot E_k[i]$ which again results in
the same score of $642.2393504256$ as above.

%%%%%%%%%%%%%%%%%%%%%%%%%%%%%%%%%%%%%%%%%%%%%%%%%%%%%%%%%%%%%%%%%%%%%%%%%%%%%
\begin{table}[tb]
\centering
\begin{tabular}{r|rrrrrrrrrr}
$E_j[i]$ & $i=1$ & 2 & 3 & 4 & 5 & 6 & 7 & 8 & 9 & 10 \\

\hline

$j=10$ &
6.720 &
7.868 &
8.730 &
9.466 &
10.160 &
10.840 &
11.534 &
12.270 &
13.132 &
14.280 \\

9 &
6.870 &
8.056 &
8.965 &
9.760 &
10.500 &
11.240 &
12.035 &
12.944 &
14.130 \\

8 &
7.038 &
8.287 &
9.241 &
10.089 &
10.911 &
11.759 &
12.713 &
13.962 \\

7 &
7.239 &
8.553 &
9.570 &
10.500 &
11.430 &
12.447 &
13.761 \\

6 &
7.479 &
8.861 &
9.970 &
11.030 &
12.139 &
13.521 \\

5 &
7.765 &
9.254 &
10.500 &
11.746 &
13.235 \\

4 &
8.120 &
9.771 &
11.229 &
12.880 \\

3 &
8.599 &
10.500 &
12.401 \\

2 &
9.292 &
11.708 \\

1 &
10.500
\end{tabular}

\medskip
\caption{The expected values $E_j[i]$ necessary to decide upon any optimal
  move in \tte.  The first row ($j=10$) is not needed to play the game but
  only to compute the expected score of the strategy.}
\label{tab:E-values}
\end{table}
%%%%%%%%%%%%%%%%%%%%%%%%%%%%%%%%%%%%%%%%%%%%%%%%%%%%%%%%%%%%%%%%%%%%%%%%%%%%%

We have depicted the complete table $E_j[i]$ for $j=1,\dots,10$ and
$i=1,\dots,j$ in Table~\ref{tab:E-values}.  This table allows us to play the
complete game using the ``maximum expected score'' strategy: Assume that
there are $j+1$ free slots and we have to place a roll of~$x$.  Find $i \in
\{1,\dots,j+1\}$ such that $E_j[i-1] \le x \le E_j[i]$.  (Recall that we assume
$E_j[0] = -\infty$ and $E_j[j+1] = +\infty$.)  Place $x$ into the $i$-th
\emph{free} slot, sorted from smallest to largest multiplier.

As an example, assume that half of the slots have been filled, so $j+1 = 5$.
Row $E_4$ from Table~\ref{tab:E-values} tells us that rolls $3$ to $8$ will
be placed into the first free slot with smallest multiplier; roll $9$ is
placed into the second free slot; rolls $10$ and $11$ are placed into the
third free slot; roll $12$ is placed into the fourth free slot; and, finally,
rolls $13$ to $18$ are placed into the last free slot with highest
multiplier.

%%%%%%%%%%%%%%%%%%%%%%%%%%%%%%%%%%%%%%%%%%%%%%%%%%%%%%%%%%%%%%%%%%%%%%%%%%%%%
%%%%  SECTION  %%%%%%%%%%%%%%%%%%%%%%%%%%%%%%%%%%%%%%%%%%%%%%%%%%%%%%%%%%%%%%
%%%%%%%%%%%%%%%%%%%%%%%%%%%%%%%%%%%%%%%%%%%%%%%%%%%%%%%%%%%%%%%%%%%%%%%%%%%%%

\section{Knowing the future}

The maximum expected score that we can reach, is significantly higher than
the score of the random strategy, but not to an extend that one might
initially think.  In particular, the maximum expected score of $642.2$ is
much smaller than the highest score of~$990$.  But the highest score can only
be reached if we have ten rolls of $18$, and the chances that this is going
to happen are
\[
  1 / 221073919720733357899776 = 4.52 \cdot 10^{-24} .
\]
For all other \tte instances, the highest score is naturally unreachable.
But with the same probability, we have ten rolls of $3$, and any strategy
will result in the minimum score of $165$.

A better way of evaluating the performance of our strategy, is to compare it
against an ``all-knowing'' strategy which is allowed to look into the future:
To this end, assume that our strategy knows the outcome of all ten rolls
before having to place the first roll.  This ``all-knowing'' strategy will
simply sort all rolls and then place them accordingly.

Again, we cannot judge the performance of this strategy by evaluating a
single game.  Instead, we play many games and sum up the scores; this again
boils down to the expected score of the strategy.  This can be computed using
``classical'' dynamic programming; we do not have to take into account the
set of slots that have been filled so far.  Let $L := \abs{B}$.  We define
$Q[y,l]$ as the partial score obtained by the ``all-knowing'' strategy for
placing $L-l$ rolls $x \ge y$, whereas for the remaining $l$ rolls we know $x
< y$ but these have not been scored so far (that is, $l$ free slots).  Then,
$Q[\xmax,L]$ is the expected score of the ``all-knowing'' strategy.  We infer
the recurrence
\begin{equation}
  Q[y,l] = \sum_{k=0,\dots,l} \binom{l}{k} p_y^k (1-p_y)^{l-k} \cdot \Bigl( y
  \cdot S(l-k+1,l) + Q[y-1,l-k] \Bigr)
\end{equation}
where
\[
  p_y := \Prob(X=y | X \le y) = \frac{\Prob(X=y)}{\sum_{x \le y} \Prob(X=x)}
\]
and
\[
  S(i,j) := \sum_{k = i,\dots,j} k = \tfrac{1}{2} \bigl( (j+1)j - i(i-1)
  \bigr) .
\]
We initialize $Q[y,0] = 0$ for all $y = \xmin,\dots,\xmax$, and $Q[\xmin-1,l]
= 0$ for all $l=0,\dots,L$.

%%%%%%%%%%%%%%%%%%%%%%%%%%%%%%%%%%%%%%%%%%%%%%%%%%%%%%%%%%%%%%%%%%%%%%%%%%%%%
\begin{table}[tb]
\centering
\begin{tabular}{lS}
Minimum possible score & 165.0 \\

Median score of the ``random'' strategy$^*$ & 577 \\

Expected score of the ``random'' strategy & 577.5 \\

Expected score of the ``maximum expected score'' strategy & 642.23935 \\

Median score of the ``maximum expected score'' strategy$^*$ & 646 \\

Expected score of the ``all-knowing'' strategy & 652.93403 \\

Median score of the ``all-knowing'' strategy$^*$ & 654 \\

Maximum possible score & 990.0
\end{tabular}

\medskip
\caption{Different important scores for the \tte game.  $^*$Median scores
  where experimentally determined from one million runs, see
  Sec.~\ref{sec:simulations}.}
\label{tab:important-scores}
\end{table}
%%%%%%%%%%%%%%%%%%%%%%%%%%%%%%%%%%%%%%%%%%%%%%%%%%%%%%%%%%%%%%%%%%%%%%%%%%%%%

We reach an expected score of $652.93403$ for the ``all-knowing'' strategy.
Somewhat surprisingly, this expected score is not much higher than the
$642.23935$ that our strategy can reach without knowing the future: Knowing
the future only gives us an expected upper hand of about ten points.  All
scores are summarized in Table~\ref{tab:important-scores}.

%%%%%%%%%%%%%%%%%%%%%%%%%%%%%%%%%%%%%%%%%%%%%%%%%%%%%%%%%%%%%%%%%%%%%%%%%%%%%
%%%%  SECTION  %%%%%%%%%%%%%%%%%%%%%%%%%%%%%%%%%%%%%%%%%%%%%%%%%%%%%%%%%%%%%%
%%%%%%%%%%%%%%%%%%%%%%%%%%%%%%%%%%%%%%%%%%%%%%%%%%%%%%%%%%%%%%%%%%%%%%%%%%%%%

\section{Implementations and simulations} \label{sec:simulations}

%% Simulation study
%% 1000000 runs
%% Average performance optimal: 642.272639
%% Average performance all-knowing: 652.947393

All algorithms presented in this paper were implemented in Groovy~1.8.6 and
run on a laptop computer.  All computations were carried out with high
precision (40+~digits).  In addition, we implemented both variants of the
``maximum expected score'' strategy and the ``all-knowing'' strategy and
performed simulations.  After one million runs, the average score of the
``maximum expected score'' strategy was $642.272639$ (for both variants), and
the average score of the ``all-knowing'' strategy was $652.947393$.  This
agrees well with the theoretical values computed above.  Running times of our
computations were negligible.  We have also computed median scores from these
evaluations, see again Table~\ref{tab:important-scores}.

%%%%%%%%%%%%%%%%%%%%%%%%%%%%%%%%%%%%%%%%%%%%%%%%%%%%%%%%%%%%%%%%%%%%%%%%%%%%%
\begin{figure}[tb]
\centering
\includegraphics[width=0.9\textwidth, trim=55 310 55 350, clip]%
  {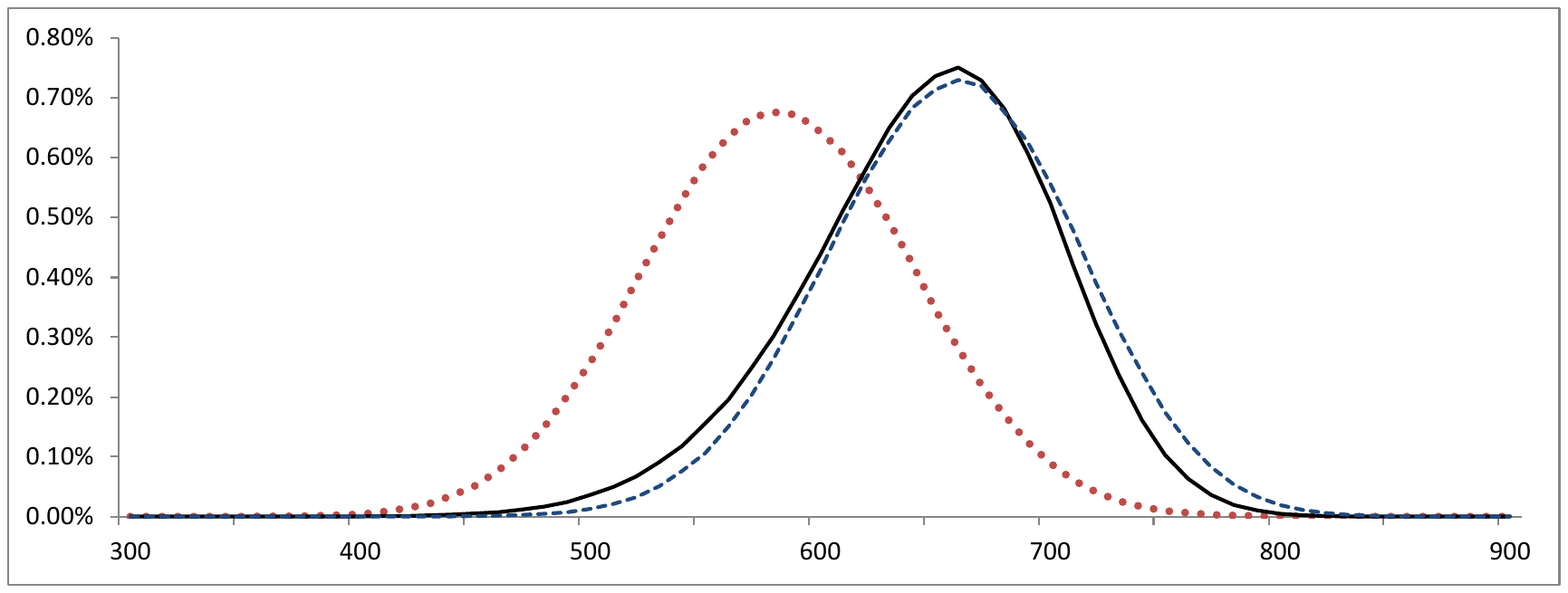}
\caption{Empirical score distribution for the random strategy (dotted line),
  the ``maximum expected score'' strategy (solid line), and the
  ``all-knowing'' strategy (dashed line).  Calculated from one million runs,
  binned using bin width~10.}
\label{fig:score-dist}
\end{figure}
%%%%%%%%%%%%%%%%%%%%%%%%%%%%%%%%%%%%%%%%%%%%%%%%%%%%%%%%%%%%%%%%%%%%%%%%%%%%%

The empirical distributions of scores are depicted in
Fig.~\ref{fig:score-dist}.  We have smoothed the curves by binning ten values
in each bin, $\{10n,\dots,10n+9\}$ for $n=15,\dots,99$.

%%%%%%%%%%%%%%%%%%%%%%%%%%%%%%%%%%%%%%%%%%%%%%%%%%%%%%%%%%%%%%%%%%%%%%%%%%%%%
%%%%  SECTION  %%%%%%%%%%%%%%%%%%%%%%%%%%%%%%%%%%%%%%%%%%%%%%%%%%%%%%%%%%%%%%
%%%%%%%%%%%%%%%%%%%%%%%%%%%%%%%%%%%%%%%%%%%%%%%%%%%%%%%%%%%%%%%%%%%%%%%%%%%%%

\section{Variants of \tte}

We have noted above that our computations are not limited to the \tte variant
where three six-sided dice are thrown.  To exemplify this claim, let us
consider one more variant, namely throwing two ``slightly loaded''
twelve-sided dice: For each die, the probability for a roll of ``12'' is
$\frac{2}{13}$, and the probability of all other rolls is~$\frac{1}{13}$.
The probabilities for throwing two loaded, twelve-sided dice are depicted in
Table~\ref{tab:probs-loaded}.  The expected value of a single roll is
$13.84615$.

%%%%%%%%%%%%%%%%%%%%%%%%%%%%%%%%%%%%%%%%%%%%%%%%%%%%%%%%%%%%%%%%%%%%%%%%%%%%%
\begin{table}[tb]
\centering
\begin{footnotesize}
\setlength\tabcolsep{4pt}
\begin{tabular}{r|rrrrrrrrrrrrrrrrrrrrrrr}
  $x$ & 2 & 3 & 4 & 5 & 6 & 7 & 8 & 9 & 10 & 11 & 12 & 13 & 14 & 15 & 16 & 17
  & 18 & 19 & 20 & 21 & 22 & 23 & 24 \\

  \hline

  $169 \cdot \Prob(X=x)$ & 1 & 2 & 3 & 4 & 5 & 6 & 7 & 8 & 9 & 10 & 11 & 14 &
  13 & 12 & 11 & 10 & 9 & 8 & 7 & 6 & 5 & 4 & 4
\end{tabular}
\end{footnotesize}

\medskip
\caption{Probabilities for throwing two loaded, twelve-sided dice.}
\label{tab:probs-loaded}
\end{table}
%%%%%%%%%%%%%%%%%%%%%%%%%%%%%%%%%%%%%%%%%%%%%%%%%%%%%%%%%%%%%%%%%%%%%%%%%%%%%

Again, we can calculate the table of expected scores for all positions, see
Table~\ref{tab:E-values-loaded}.  Assuming slot multipliers $1$ to $5$, we
reach a score of $231.11229$ for the ``maximum expected score'' strategy, and
$236.97840$ for the ``all-knowing'' strategy.  In comparison, the random
strategy reaches an expected score of $207.69231$.

%%%%%%%%%%%%%%%%%%%%%%%%%%%%%%%%%%%%%%%%%%%%%%%%%%%%%%%%%%%%%%%%%%%%%%%%%%%%%
\begin{table}[tb]
\centering
\begin{tabular}{r|rrrrr}
$E_j[i]$ & $i=1$ & 2 & 3 & 4 & 5 \\

\hline

$j=5$ &
9.038 &
11.713 &
13.868 &
16.012 &
18.599 \\

4 &
9.680 &
12.613 &
15.113 &
17.978 \\

3 &
10.532 &
13.861 &
17.146 \\

2 &
11.753 &
15.939 \\

1 &
13.846
\end{tabular}

\medskip
\caption{The expected values $E_j[i]$ necessary to decide upon any optimal
  move in the game with two loaded twelve-sided dice.  The first row ($j=5$)
  is not needed to play the game but only to compute the expected score of
  the strategy.}
\label{tab:E-values-loaded}
\end{table}
%%%%%%%%%%%%%%%%%%%%%%%%%%%%%%%%%%%%%%%%%%%%%%%%%%%%%%%%%%%%%%%%%%%%%%%%%%%%%

%%%%%%%%%%%%%%%%%%%%%%%%%%%%%%%%%%%%%%%%%%%%%%%%%%%%%%%%%%%%%%%%%%%%%%%%%%%%%
%%%%  SECTION  %%%%%%%%%%%%%%%%%%%%%%%%%%%%%%%%%%%%%%%%%%%%%%%%%%%%%%%%%%%%%%
%%%%%%%%%%%%%%%%%%%%%%%%%%%%%%%%%%%%%%%%%%%%%%%%%%%%%%%%%%%%%%%%%%%%%%%%%%%%%

\section{Conclusion}

We have presented the game \tte, plus a strategy to maximize the expected
score.  In addition, we have shown how to compute the expected score of an
omniscient strategy.

Playing the strategy maximizing the expected score, does not maximize your
chances to win a two-player game: That is, two players are given the same
rolls and compete against each other to maximize the score reached in a
single game.  The player that wins the most games wins the match.  Again, we
assume that a sufficiently large number of games is played.  Here, the
``maximum expected score'' strategy introduced in this paper will be hard to
beat.  But if you know your opponent is playing this strategy, then you can
still get an upper hand against the score-optimal strategy: It is enough to
be a few points ahead in most games, whereas the score of any lost game is
unimportant.  This takes us into the realms of game theory; in particular,
there is no longer one optimal strategy but instead, there may be cases where
strategy A beats strategy B, B beats C, but C beats~A.  Things will become
even more complicated in multi-player games where several strategies compete
simultaneously.  But at least, it should be possible to come up with a
strategy that beats the score-optimal strategy in a two-player game: we only
have to consider the chances of some score being higher than that of the
score-optimal strategy in every move.  The state space will increase
considerably, because now we have to consider the filled slots and the score
obtained so far.

%%%%%%%%%%%%%%%%%%%%%%%%%%%%%%%%%%%%%%%%%%%%%%%%%%%%%%%%%%%%%%%%%%%%%%%%%%%%%
%%%%  BIBLIOGRAPHY ETC  %%%%%%%%%%%%%%%%%%%%%%%%%%%%%%%%%%%%%%%%%%%%%%%%%%%%%
%%%%%%%%%%%%%%%%%%%%%%%%%%%%%%%%%%%%%%%%%%%%%%%%%%%%%%%%%%%%%%%%%%%%%%%%%%%%%

\paragraph{Acknowledgments.}

I want to thank my school teacher, Klaus Bovermann, for introducing this game
to us, and getting us into developing strategies for it.

\end{document}